\documentclass[aps,epsf,preprint]{revtex4}
\usepackage{amssymb}
\usepackage{amsmath}
\usepackage{color}

\def\[{\left[}
\def\]{\right]}
\def\be{\begin{eqnarray}}
\def\ee{\end{eqnarray}}
\def\bm{\begin{pmatrix}}
\def\em{\end{pmatrix}}
\def\ba{\begin{array}}
\def\ea{\end{array}}
\def\bi{\begin{itemize}}
\def\ei{\end{itemize}}

\def\({\left(}
\def\){\right)}

\def\eq#1{Eq.(\ref{#1})}
\def\a{\alpha}

\def\m{\mu}

\def\n{\nu}

\def\labels#1{\label{#1}}
\def\bn{\begin{enumerate}}

\def\en{\end{enumerate}}
\def\b{\beta}
\def\g{\gamma}
\def\ba{\begin{array}}
\def\ea{\end{array}}
\def\bc{\begin{center}}
\def\ec{\end{center}}

\def\.{\!\cdot\!}
\def\igw#1{\includegraphics[width=#1cm]}
\def\igwg#1#2{\igw{#1}{#2.png}} 
\def\+{\!+\!}
\def\-{\!-\!}

\def\r{\rho}
\def\h{{1\over 2}}

\def\={\stackrel{.}{=}}

\def\bn{\bar n}
\def\1{\bar 1}
\def\8{\bar 8}
\def\9{\bar 9}
\def\7{\bar 7}

\usepackage{amssymb}
\usepackage{graphicx}

\begin{document}
\title{A Model of the Black Hole Interior}
\author{C.S. Lam}
\affiliation{Department of Physics, McGill University\\
 Montreal, Q.C., Canada H3A 2T8\\
Department of Physics and Astronomy, University of British Columbia,  Vancouver, BC, Canada V6T 1Z1 \\
Email: Lam@physics.mcgill.ca}

\begin{abstract}
A model is proposed for the interior of a neutral non-rotating black hole. It consists of an ideal fluid with density 
$\r$ and a negative pressure $p$, obeying the equation of state $p=-\xi\r$. In order 
for the Einstein equation to have a solution, $\xi$
must lie in a narrow range between  0.1429  and 0.1716.
\end{abstract}

\maketitle

\section{Introduction}
The interior of a black hole is a mystery. When a star collapses into a black hole, its matter loses the
baryonic number, fermionic number, isotopic spin, hypercharge etc.,
to become a different kind of matter that shall be referred to as black hole matter. The nature of this new  matter 
is largely unknown, and its distribution
inside the black hole is also unknown, except that there is a singularity at the center  \cite{aK56, aR57, rP65, SG14, kL22} which can likely be washed out when quantum effect is taken into account \cite{mB07, aP17, ABP19, AQS23}. 

The property of this new matter is of considerable interest by itself. Moreover, its quantization might lead to a new understanding of the Hawking radiation. It might even be related to the inflaton matter before reheating because the latter presumably
also carries no global quantum numbers. However,
being inside the black hole, this new  matter cannot be  experimentally measured, so we must resort to theoretical modelling
guided and constraint by the Einstein equation to gain some understanding of it.

In this article we propose a simple model of black-hole matter, described by an ideal fluid with a positive mass density $\r$,
and an equation of state $p=-\xi\r$ with a negative pressure. To prevent matter from sinking to the center,
a repulsive force ($\xi>0$) produced by a negative pressure is required. In order to have a solution,  it turns out that only a very narrow range of $\xi$ between 0.1429  and 0.1716 is allowed. 
In this way the present model differs from those \cite{MM15, BMN19, MM23} where black-hole matter is made up of dark energy with $\xi=1$.
The density
and the metric of this model have a singularity at the origin as per the singularity theorem, but the internal mass in a finite volume
is finite and  vanishes as the volume goes to zero.

\section{The Model}
We would like to construct a model of a spherical black hole 
whose interior space is filled with matter. Assuming there is nothing else in spacetime, the stress-energy-momentum tensor $T_{\m\n}$ is zero outside the black hole,
but it is non-zero inside where matter is. Black hole matter is assumed to be an ideal fluid with
\be T_{\m\n}=(p+\r)U_\m U_\n+p g_{\m\n}, \labels{semt}\ee
whose four-velocity is as usual normalized to $U^\m U_\m=-1$. The energy density $\r$ is assumed to be positive, but 
the pressure determined by the equation of state  $p=-\xi\r$ can be either positive ($\xi<0$) or negative ($\xi>0$) at this point. 

The metric outside such a black hole with mass $M$ is the Schwarzschild metric
\be ds^2&=&-\bigg(1-{2GM\over r}\bigg)dt^2+\bigg(1-{2GM\over r}\bigg)^{-1}dr^2+r^2 d\Omega^2.\ee
It obeys the Einstein equation $G_{\m\n}=8\pi GT_{\m\n}=0$, with the event horizon
located at $r=R=2GM$.

The metric inside is
given by the line element \cite{sC04}
\be ds^2&=&-e^{2\a(r)}dt^2+\bigg(1-{2Gm(r)\over r}\bigg)^{-1}dr^2+r^2 d\Omega^2.\labels{ds2}\ee
It obeys the Einstein equation $G_{\m\n}=8\pi GT_{\m\n}$, with  $T_{\m\n}$ given by 
\eq{semt}.

With this energy-momentum tensor,  Einstein equation is equivalent to the Tolman-Oppenheimer-Volkoff equation
for the pressure gradient  \cite{sC04, OV39} 
\be {dp\over dr}&=&-{(\r+p)[Gm(r)+4\pi Gr^3p]\over r[r-2Gm(r)]},\qquad {\rm where}\labels{dpdr}\\
m(r)&=&4\pi\int_0^r\r(r'){r'}^2dr'\labels{massr}\ee
is the mass of a black-hole matter ball of radius $r\le R$.
The  metric functions $\a$ obeys \cite{sC04}
\be {d\a\over dr}&=&-{1\over (\r+p)}{dp\over dr}.\labels{dadr}\ee

To match the inside of the black hole to the outside, we need to specify 
boundary conditions at $r=R$. Since $T_{\m\n}=0$ outside but non-zero inside, it
cannot be analytic at $r=R$, for otherwise it has to vanish inside as well. 
As a result, the functions specifying the metric cannot be analytic either.
Instead, we require continuity at the event horizon, namely,
\be m(R)=M,\quad p(R)=0,\quad \r(R)=0, \quad e^{2\a(R)}=0,\labels{bondc}\ee
and these conditions are sufficient for the Einstein equation inside the event horizon to have a solution.
 
For $R-r$ small and positive, \eq{dpdr}
can be approximated by
\be {dp\over dr}=-{(\r+p)GM\over R[r-2GM]}={\xi-1\over 2\xi}{p\over R-r},\quad (R-r\ll R),\labels{dpdrR}\ee
whose solution for small and positive $R-r$ is
\be p(r)\simeq -\tilde c(R-r)^{(1-\xi)/2\xi}\equiv -\tilde c(R-r)^\g \labels{pR}\ee
for some $\tilde c>0$. In order for $p(R)=0$, it is necessary to have $0<\xi<1$, so the black-hole matter necessarily carries
a negative pressure. 

Using \eq{dadr} and $p=-\xi\r$, one gets 
\be e^{2\a(r)}=c'\r(r)^{1/\g}\labels{e2a}\ee
for some $c'>0$. Thus $e^{2\a(r)}$ is positive throughout the interior of the black hole. In particular, when $r$ approaches
$R$, \eq{pR} shows that $e^{2\a(r)}\to c''(R-r)$ for some $c''>0$,  satisfying
the boundary condition in \eq{bondc}.

It would be simpler to write the Tolman-Oppenheimer-Volkoff equation in a dimensionless form. To that end,
let $x=r/R,\ \bar m(x)=m(r)/M$, and $\bar \r(x)=\r(r)(R^3/M)$. Then \eq{dpdr} can be written as
\be
{d\bar \r(x)\over dx}&=&-{\xi-1\over 2\xi}{\bar \r(x)\over x}{\bar m(x)-4\pi\xi x^3\bar \r(x)\over x-\bar m(x)},\labels{dless1}\\
{d\bar m(x)\over dx}&=&4\pi x^2 \bar \r(x), \labels{dless2}\ee
with the boundary condition $\bar\r(1)=0$ and $\bar m(1)=1$. The interior solution equivalent to \eq{pR} near $x=1$ is then
\be
\bar \r(x)&\simeq& c(1-x)^{(1-\xi)/2\xi},\\
\bar m(x)&\simeq&1-{8\pi \xi c\over \xi+1}(1-x)^{(\xi+1)/2\xi},\qquad (1-x\ll 1)\label{solrm}\ee
for some $c>0$. 

Let us turn to the behavior near $x=0$. Assuming the mass function $\bar m(x)$ to increase
steadily from  0 to 1 in the interval $x\in(0,1)$, then for small $x$, $\bar m(x)=\m_0x^\b$ for some $\b>0$ and $\m_0>0$. It follows from \eq{dless1} that $\bar \r(x)=(\m_0\b/4\pi) x^{\b-3}$, so $x^3\bar \r(x)$ is proportional to $\bar m(x)$,
and  $\bar m(x)-4\pi\xi x^3\bar \r(x)=\bar m(x)(1-\b\xi)$.

If $\b>1$, then $x-\bar m(x)\simeq x$ for sufficiently small $x$, in which case the $x$-behavior
on both sides of \eq{dless1} can never match, so in order to have a solution,
we need to have $\b\le 1$. 
As a result, $\bar\r(x)\sim x^{\b-3}$ must diverge near $x=0$. \eq{dadr}  shows that  $e^{2\a}\simeq (\bar\r/\bar\r_0)^{2\xi/(1-\xi)}$, which also diverges near $x=0$, but it remains positive throughout
the interior of the black hole. To avoid another singularity between $x=0$ and $x=1$,
we must keep $2Gm(r)>r$, or $\bar m(x)>x$ in this range.
Hence the signature of the metric changes from $(-,+,+,+)$ outside the black hole to
$(-,-,+,+)$ inside, whereas the signature $(+,-,+,+)$ of the Schwarzschild metric   inside is that of a Kantowski-Sachs metric.

The presence of a singularity is hardly surprising \cite{aK56, aR57, rP65, kL22, SG14}. Using the Raydhuchaudri equation, it can be shown that a singularity is present at $r=0$ if the convergence condition
 $R_{\m\n}U^\m U^\n\ge 0$ is satisfied. In the present case, the Einstein equation
$R_{\m\n}-\h R g_{\m\n}=8\pi G T_{\m\n}$ implies 
$R_{\m\n}=8\pi G(T_{\m\n}-\h T^a_\a g_{\m\n})=8\pi G\[(p+\r)U_\m U_\n+\h(\r-p)g_{\m\n}\]$, hence
$R_{\m\n}U^\m U^\n=4\pi G\r(1-3\xi)$. With $\xi<1/3$, as shall be presently shown, the 
convergence condition is met so a singularity is present.

If $\b<1$, then $x-\bar m(x)\simeq -\bar m(x)$ for sufficiently small $x$, so \eq{dless1} is satisfied when
\be \b=(\-1 \+ 7 \xi)/\xi (1 \+ \xi).\labels{beta}\ee
In order for $\b$ to be between 0 and 1, $\xi$ is allowed only a narrow range of values between 
$\xi=1/7\simeq 0.1429$ (when $\b=0$) and  $\xi=3-2\sqrt{2}\simeq 0.1716$ (when $\b=1$).
For larger values of $\xi$, this formula yields $\b>1$, so it is not allowed.

A numerical solution for $\xi=0.16$, somewhere in the middle of the allowed range, is shown in Fig.~1 as an illustration. 
This solution is obtained numerically by integrating \eq{dpdr} and \eq{massr}, starting at an initial $x$ so small that the approximation  $\bar m(x)=\m_0x^\b$ and $\bar \r(x)=(\m_0\b/4\pi) x^{\b-3}$ is accurate. The
value of $\b$ given by \eq{beta}, and the constant $\m_0$ is adjusted to yield $\bar m(1)=1$ and $\bar\r(1)=0$ at the
other boundary.

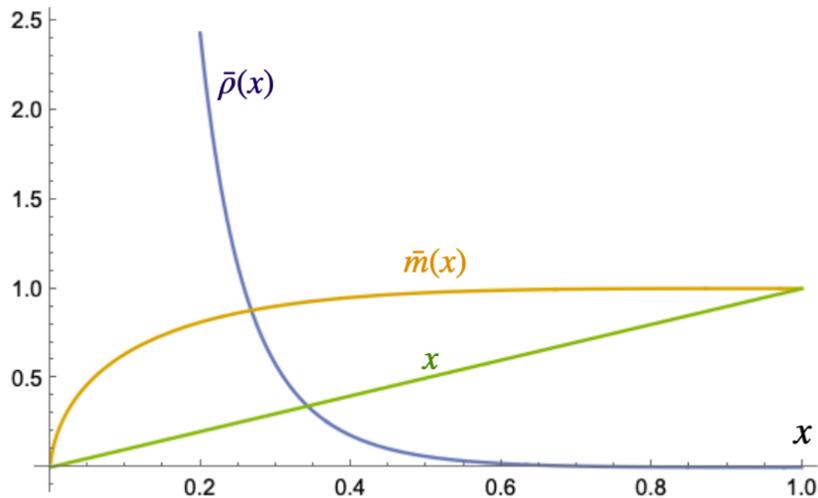
\bc\igwg{12}{Fig1}\\ Fig.~1.\quad Scaled mass and density distribution in the black-hole interior for $\xi=0.16$\ec

I am grateful to Bei-Lok Hu for discussions and suggestions.

\end{document}